\documentclass[11pt]{article}
\usepackage{blois2002,epsfig,axodraw}
\def\be{\begin{equation}}
\def\ee{\end{equation}}
\def\bea{\begin{eqnarray}}
\def\eea{\end{eqnarray}}

\begin{document}
\vspace*{4cm}
\title{TESTS OF LEPTOGENESIS AT LOW ENERGY}

\author{ T. HAMBYE }

\address{Centre de Physique Th\'eorique, Luminy Case 907, 
13288 Marseille Cedex 09,
France\\Scuola Normale Superiore, Piazza dei Cavalieri 7, 56126 Pisa, Italy}

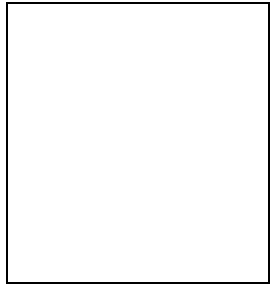
\maketitle\abstracts{
The problem of testing leptogenesis from low energy experiments is discussed
following three different perspectives.
Firstly, we review the prospects that from low energy experiments 
we could reconstruct the neutrino Yukawa 
coupling matrix and hence constrain the leptogenesis mechanism.
We emphasize the fact that the experimental determination of
the phases and mixings 
in the light neutrino mass matrix is irrelevant for leptogenesis, 
unless additional information about the texture of the Yukawa coupling matrix
is provided by other observables.
Secondly, we show how the discovery of an extra gauge boson could bring 
us important indications for leptogenesis. Thirdly, we discuss the problems 
one encounters when attempting to build a leptogenesis mechanism at a 
directly testable scale, presenting an explicit model which avoids
these problems. 
}

\section{Introduction}

The matter-antimatter asymmetry of the universe is one of the most 
fascinating enigma of today's particle physics and cosmology.
The recent evidence for small but non-vanishing neutrino masses has established
the leptogenesis mechanism with heavy singlet 
neutrinos\cite{FY} as one of the most 
appealing explanation of 
this asymmetry.
In the seesaw mechanism\cite{Seesaw} with 
heavy singlet neutrinos, the smallness of 
neutrino masses is naturally explained and the terms at their origin,
(i.e. the Yukawa interactions involving the left-handed leptons 
and the singlet neutrinos, 
and the lepton number violating singlet neutrino Majorana mass term),
are also expected to be at the origin of a cosmological 
lepton asymmetry produced at 
the epoch of 
the singlet neutrino decays. 
Once produced, the lepton asymmetry $Y_L$ 
is expected to be converted for a large
fraction to a baryon asymmetry $Y_B$ 
by the effects of the sphalerons\cite{Sphaler}
associated to the B+L anomaly
\begin{equation}
Y_B=-C Y_L \, ,
\end{equation}
with $Y_{B,L}=n_{B,L}/s$, the baryon (lepton) number density 
over the entropy density. 
In the seesaw extended standard model 
and seesaw extended minimal supersymmetric
standard model, $C$ is equal to $28/79$ and $8/23$ respectively. From
nucleosynthesis constraints, $Y_B$ has been determined to be 
within the range\cite{YBrange},
$3 \cdot 10^{-11} < Y_B < 9 \cdot 10^{-11}$, in agreement with recent
results from CMB data.
Qualitatively if the singlet neutrinos have masses between
$\sim 10^{10}$~GeV
and the GUT scale, the typical values of the Yukawa couplings we need 
for inducing the leptogenesis with the right order of magnitude are also the 
ones we need for inducing neutrino masses in agreement with the 
neutrino experiments 
and with the dark matter bound $m_\nu < \hbox{few eV}$.
Moreover for such values of the singlet neutrino masses, the 
seesaw leptogenesis model can be embedded naturally in grand unified models
such as the ones based on 
SO(10) which predicts the existence of singlet neutrinos.

For all these reasons this seesaw leptogenesis model is definitely 
very attractive. More pragmatically, however, this model has a major 
default: it's very difficult to test it. With so large singlet 
neutrino masses, it is not possible to test it directly 
by producing those particles.
The problem of how to test leptogenesis indirectly from low energy observables
is therefore crucial.
In these proceedings we will discuss this problem following three different 
perspectives.
First, in section 2 the various observables from which we could 
build a program of low-energy tests of leptogenesis are discussed.
The question we address is how we could 
reconstruct the Yukawa coupling matrix from these observables
and from that how we could test leptogenesis.
Secondly, in section 3, we emphasize the fact that the discovery 
of an extra gauge boson around the 
TeV scale would bring very interesting informations about leptogenesis, 
and discuss this question in particular in the context of a unified theory 
based on 
the $E_6$ group\cite{E6}. 
Finally in section 4 we discuss the problems one encounters if, instead
of considering leptogenesis models at a very high scale, we want to build
a leptogenesis mechanism at a directly testable scale. We present 
an explicit leptogenesis model at the TeV scale which avoids 
these problems \cite{THTeV}.

\section{Low energy observables and leptogenesis}

The seesaw mechanism is usually 
implemented in a minimal way by adding
to the standard model (SM) three heavy singlet neutrinos $N_i$ coupling to the
left-handed lepton doublet $L$ through Yukawa interactions:
\begin{equation}
{\cal L} = {\cal L}_{SM} + N_i^c (Y_\nu)_{ij} L_j H -\frac{1}{2} N_i^c 
(M_N)_{ij} N_j^c \label{Lagr} \, ,
\end{equation}
Without loss of generality, one can always 
work in the basis where the charged lepton and heavy 
singlet neutrino mass matrices are real and diagonal.
In this basis we have also the freedom to redefine the three
doublets of left-handed 
lepton fields
by multiplying them by 
any phase without affecting the physical content of the model. In 
full generality 
the lagrangian above involves therefore 
9 real parameters and 6 phases in $Y_\nu$
in addition to the 3 real singlet neutrino masses $M_{N_i}$.
The light neutrino masses induced are given by the usual seesaw formula:
\begin{equation}
{\cal M}_\nu = Y_\nu^T (M_N)^{-1} Y_\nu v^2 \label{Mnu} \, ,
\end{equation}
with $v=174$~GeV.
Being symmetric this matrix can be diagonalized by a unitary matrix $U$,
\begin{equation}
U^T {\cal M}_\nu U = {\cal M}_\nu^D \label{MnuD} \, ,
\end{equation}
with ${\cal M}_\nu^D=diag(m_{\nu_1},m_{\nu_2},m_{\nu_3})$. $U$ 
can be written as $U=PV$ where $V$ is a CKM type matrix involving 3 angles 
and one phase $\delta$,
\begin{equation}
V = \pmatrix{ c_1 c_3 & - s_1 c_3 & -s_3 e^{- i \delta} \cr
s_1 c_2 - c_1 s_2 s_3 e^{ i \delta} & c_1 c_2 + s_1 s_2 s_3 e^{ i \delta} &
- c_3 s_2 \cr s_1 s_2 + c_1 c_2 s_3 e^{ i \delta} &
c_1 s_2 - s_2 c_2 s_3 e^{ i \delta} & c_2 c_3 }\, ,
\end{equation}
and where the matrix $P \equiv diag(e^{i \phi_1}, e^{i \phi_2},1)$ involves 
two Majorana phases $\phi_{1,2}$ which cannot be absorbed in the $N_i$ fields
(i.e. without reappearing in the $M_{N_i}$'s).

In the thermal history of the universe, at temperature around $T\sim M_{N_i}$,
the $N_i$'s 
disappear decaying to left-handed leptons and scalar Higgs bosons.
The averaged $\Delta L$ produced per decay is given by the CP-asymmetry:
\begin{equation}
\varepsilon_i=\frac{\Gamma(N_i \rightarrow l  H) -\Gamma(N_i \rightarrow
\bar{l} H^\ast)}{
\Gamma(N_i \rightarrow l   H) +\Gamma(N_i \rightarrow
\bar{l} H^\ast)} \, . \label{epsilonidef}
\end{equation}
At lowest order, $\varepsilon_i$ is given by the interference of the 
tree and one loop (vertex\cite{FY} and self-energy\cite{Flanz}) diagrams 
which gives:
\begin{equation}
\varepsilon_i=-\frac{1}{8 \pi} \sum_l \frac{\mbox{Im}
\left[ \left( Y_\nu Y_\nu^\dagger \right)^{il} 
\left(Y_\nu Y_\nu^\dagger\right)^{il} \right]}{\sum_j
|Y_\nu^{ij} |^2} \sqrt{x_l} \left[ \log (1+1/x_l)+\frac{2}{x_l-1} \right] \, ,
\label{epsiloniresult}
\end{equation}
with $x_l=(M_{N_l}/M_{N_i})^2$.
If the out-of-equilibrium decay condition\cite{Sakharov}
\begin{equation}
\Gamma_{N_i}=\frac{1}{8 \pi} \sum_j |Y_\nu^{ij}|^2 M_{N_i} < H(T=M_{N_i})=
\sqrt{\frac{4 \pi^3 g_\ast}{45}} \frac{T^2}{M_{Planck}}\Big|_{T=M_{N_i}} \, ,
\label{ooec}
\end{equation}
is satisfied, then the lepton asymmetry density will be given by
$Y_L \sim \sum \varepsilon_i/g^\ast$ with $g^\ast$ the number of 
active degrees of freedom at this temperature ($g_\ast \sim 100$).

In order to determine the constraints on this leptogenesis mechanism
we could obtain from
low energy experiments, it is necessary to know how we could reconstruct 
the $Y_\nu$ matrix from the knowledge we have on the light neutrino 
mass matrix.
To this end, it is convenient to use the parametrization\cite{CaIb}
\begin{equation}
Y_\nu=\frac{1}{v}(M_N)^{1/2} R ({\cal M}^D_\nu)^{1/2} U^{-1} \, ,
\label{Rdef}
\end{equation}
with $R$ a complex matrix which from Eqs.~(\ref{Mnu}) and
(\ref{MnuD}) turns out to be 
orthogonal ($R^T R=1$).
This parametrization is interesting because by construction, putting 
Eq.~(\ref{Rdef}) in 
Eq.~(\ref{Mnu}), 
the low energy neutrino mass matrix
${\cal M}_\nu$ is independent of $R$, and by taking the full set 
of possible complex orthogonal matrices $R$, we can determine 
the full set of possible matrices $Y_\nu$ which give rise to the same low
energy neutrino mass matrix ${\cal M}_\nu$. Therefore $R$ contains 
all the information in $Y_\nu$ which is not contained in ${\cal M}_\nu$
and is independent of it. 
Since a general complex orthogonal matrix can be parametrized in terms of 3
real angles and 3 phases, this information depends on 6 parameters 
as it should be (i.e. 15 in $Y_\nu$ minus 9 in ${\cal M}_\nu$).
A convenient way to parametrize a complex orthogonal matrix is in terms of
three complex angles:
\begin{equation}
R = \pm \pmatrix{ \hat{c}_2 \hat{c}_3 & - \hat{c}_1 \hat{s}_3 
- \hat{s}_1 \hat{s}_2 \hat{c}_3 & \hat{s}_1 \hat{s}_3 - \hat{c}_1 \hat{s}_2
\hat{c}_3 \cr
\hat{c}_2 \hat{s}_3  & \hat{c}_1 \hat{c}_3 -\hat{s}_1 \hat{s}_2 \hat{s}_3 &
- \hat{s}_1 \hat{c}_3 - \hat{c}_1 \hat{s}_2 \hat{s}_3 \cr 
\hat{s}_2 & \hat{s}_1 \hat{c}_2 & \hat{c}_1 \hat{c}_2 }\, ,
\label{complexangles}
\end{equation}
with the ``$\pm$'' in front of the matrix to account for possible reflections.

Having the full set of possible $Y_\nu$ matrices for a given low 
energy matrix ${\cal M}_\nu$ one can now address the question:
which constraints  ${\cal M}_\nu$ gives on leptogenesis?
In Eq.~(\ref{epsiloniresult}) the lepton 
asymmetry depends only on the $M_{N_i}$ and 
on the combination $Y_\nu Y^\dagger_\nu$ which from Eq.~(\ref{Rdef}) is:
\begin{equation}
Y_\nu Y^\dagger_\nu = \frac{1}{v^2}
(M_N)^{1/2} R ({\cal M}_\nu^D) R^\dagger (M_N)^{1/2}
\label{YnuYnuD}
\end{equation}
This combination depends on the light and heavy 
neutrino mass eigenvalues as well as on $R$
but turns out to be independent of U!
This means that, in a model independent way, leptogenesis is independent
of the three mixing angles and phases in $U$ and can be expressed
in such a way that it depends only on mixing angles and phases which 
decouple from the low energy neutrino mass matrix. Of course we could 
find other parametrizations\cite{DaIb2} where 
$Y_\nu Y_\nu^\dagger$ has for example
a dependence on
the CKM phase $\delta$ in $U$ for fixed values of the other phases but 
this dependence is completely parametrization dependent and 
therefore meaningless. The fact that there exists one parametrization
(i.e.~Eq.~(\ref{Rdef})) where leptogenesis and $\delta$ are totally 
independent shows it clearly.
In other words, to speak about a "phase overlap" between the 
leptogenesis phase and 
the phase $\delta$, i.e. to know if we expect that the observation of 
a large $\delta$ phase at neutrino factories would lead naturally
to a large leptogenesis phase \cite{DaIb2}, is parametrization dependent and 
therefore arbitrary. 
Only by assuming specific textures on $Y_\nu$ relating the phases, the 
knowledge
of $\delta$ might tell us something about the leptogenesis 
phase\cite{Branco}. To assume a
specific texture on ${\cal M}_\nu$ is not sufficient because it doesn't tell
anything about the phases appearing in Eqs.~(\ref{epsiloniresult}) and
(\ref{YnuYnuD}).

To constrain leptogenesis from low energy experiments we need observables
which are sensitive to $R$. If on the one hand in the seesaw 
extended SM it is unlikely 
that some observables could give model independent 
constraints on $R$, on the other hand in the seesaw extended
minimal supersymmetric model (which involves the same number of parameters 
in $Y_\nu$), one might get a chance to reconstruct $R$ and hence 
$Y_\nu$. The slepton mass matrices, the electric dipole moments of the 
electron $d_e$ and muon $d_\mu$, the CP-conserving flavor changing 
processes $\mu \rightarrow e \gamma$ and 
$\tau \rightarrow l \gamma$, the CP-violating and CP-conserving component 
of $\mu \rightarrow eee$ and $\tau \rightarrow 3l$, 
depend on $Y_\nu$ from 
the effects of $Y_\nu$ on the renormalization (RGE's) of the 
soft supersymmetry breaking parameters they involve (i.e. the slepton doublet 
and charged singlet slepton mass terms as well as trilinear couplings,
see Ref.~\cite{CaIb,DaIb1,ElILR,ElHRS,DaIb2}). At lowest order, all 
these processes 
have the property to depend on $Y_\nu$ through an hermitian matrix 
which has the form $H= Y_\nu^\dagger D Y_\nu$ with $D$ a diagonal real matrix.
In particular the soft slepton mass matrices depend on $H$ with 
$D=\mbox{log}(M_{GUT}/M_{N_i}) \delta_{ij}$.
In Ref.~\cite{DaIb1,ElILR,ElHRS,DaIb2} it has been 
pointed out that from these observables 
and ${\cal M}_\nu$ one could reconstruct $Y_\nu$ and see the consequences 
for leptogenesis. Replacing first $v$ by $v \sin \beta$ 
in Eqs.(\ref{Mnu}), (\ref{Rdef}) and (\ref{YnuYnuD}) for the 
MSSM, the explicit way to proceed is the following\cite{ElHRS}. 
From a given 
$H$ matrix determined from 
these observables and from a given set of parameters in ${\cal M}_\nu$
one can calculate the matrix
\begin{equation}
H'=({\cal M}_\nu^D)^{-1/2} U^\dagger H U ({\cal M}_\nu^D)^{-1/2} 
v^2 \sin^2 \beta \,.
\end{equation}
The matrix $R$ as well as $(M_N)$ can then be determined as
the solution of the equation 
\begin{equation}
H'=R^\dagger (D_{ii} M_{N_i}) R \, .
\end{equation}
Thus from the nine effective low energy neutrino parameters and from the
nine parameters in the hermitian 
matrix $H$ one can determine $R$ and $(M_N)$ and hence $Y_\nu$ 
via Eq.~(\ref{Rdef}).\footnote{Note 
that, as pointed out in Ref.~\cite{ElHRS}, an hermitian matrix cannot always
be diagonalized by a complex orthogonal matrix. In this procedure
it is therefore necessary to exclude the ranges of values of $H$ which 
are for this reason not physical.}

It is worth to take few examples of matrices $H$ and ${\cal M}$
which satisfy the various experimental neutrino 
constraints as well as give an asymmetry within its experimental 
range.\cite{YBrange}
\begin{table}[t]
\begin{eqnarray*}
\begin{array}{|c||c|c|c|}
\hline
\rule{0cm}{5mm}
 \,\,\hbox{Observable} \,\, & \,\, \hbox{Current experimental} \,\, 
& \,\, \hbox{Expected experimental} \,\, & \,\, \hbox{Theoretical value} \,\,
\\ 
&  \hbox{sensitivity} &  \hbox{sensitivity} & \hbox{reachable} 
\\[0.5mm]
\hline
\hline
\rule{0cm}{5mm}
Br(\mu \rightarrow e \gamma )& \,\, < 1.2 \cdot 10^{-11}\,\, 
\, [\mbox{\cite{PDG}}]  \,\, 
& \sim 10^{-14}-10^{-15} \,\,\, [\mbox{\cite{PSI,nufact}}] & 
\sim 10^{-10} 
\,\,\, [\mbox{\cite{CaIb,ElILR,ElHRS}}] 
\\[0.5mm]
Br(\tau \rightarrow \mu \gamma) & \,\, < 6 \cdot 10^{-7} 
\,\,\, [\mbox{\cite{Inami}}]  \,\, 
& \sim 10^{-7}-10^{-8} \,\,\,[\mbox{\cite{Inami}}] & \sim 10^{-6} \,\, 
\,[\mbox{\cite{ElILR,ElHRS}}] 
\\[0.5mm]
Br(\tau \rightarrow e \gamma )& \,\, < 2.7 \cdot 10^{-6} \,\,\,
[\mbox{\cite{PDG}}]
  \,\, 
& \sim 10^{-7}-10^{-8} \,\,\, [\mbox{\cite{Inami}}] & \sim 10^{-6} \,\,
\,[\mbox{\cite{ElILR,ElHRS}}]
\\[0.5mm]
d_e \, (e \, \hbox{cm})& < 1.6 \cdot 10^{-27} \,\,\, [\mbox{\cite{Regan}}] 
& \sim 10^{-32} \,\,\,[\mbox{\cite{Lamor}}]& 
\sim 10^{-28} \,\,\,[\mbox{\cite{ElHRS}}]
\\[0.5mm]
d_\mu \, (e \, \hbox{cm})& (3.7 \pm 3.4) \cdot 10^{-19} \,\,
\,[\mbox{\cite{PDG}}] 
& \sim 10^{-24}-10^{-25} \,\,\,[\mbox{\cite{nufact,BNL}}]\, 
& \sim 10^{-25} \,\,\,[\mbox{\cite{ElHRS}}]
\\[0.5mm]
d_\tau \, (e \, \hbox{cm})& < 4.5 \cdot 10^{-17} \, \,\,[\mbox{\cite{Inami}}]&
\sim 10^{-17} \,\,\,[\mbox{\cite{Inami}}] & 
\\[0.5mm]
Br(\mu \rightarrow 3 e) &  < 1 \cdot 10^{-12}\,\, \, [\mbox{\cite{PDG}}] & 
\sim 10^{-15}-10^{-16} \,\,\,
[\mbox{\cite{PSI,nufact}}] &  \sim 10^{-12} \,\,\,[\mbox{\cite{ElILR}}] 
\\[0.5mm]
Br(\tau \rightarrow 3 e) & < (2-4) \cdot 10^{-7}\, 
\,\,[\mbox{\cite{Inami}}] &  
\sim 10^{-8} \,\,\,[\mbox{\cite{Inami}}] & 
\sim 10^{-8} \,\,\,[\mbox{\cite{ElHRS}}] 
\\[0.5mm]
Br(\tau \rightarrow \mu 2 e )& < (2-4) \cdot 10^{-7}\,\,\, 
[\mbox{\cite{Inami}}] & 
\sim 10^{-8} \,\,\,[\mbox{\cite{Inami}}] & 
\sim 10^{-8} \,\,\,[\mbox{\cite{ElHRS}}] 
\\[0.5mm]
Br(\tau \rightarrow 3 \mu) & < (2-4) \cdot 10^{-7}\,\,\, 
[\mbox{\cite{Inami}}] & 
\sim 10^{-8} \,\,\,[\mbox{\cite{Inami}}] &
\sim 10^{-9} \,\,\,[\mbox{\cite{ElHRS}}] 
\\[0.5mm]
Br(\tau \rightarrow e 2 \mu) & < (2-4) \cdot 10^{-7}\,\, \, 
[\mbox{\cite{Inami}}] & 
\sim 10^{-8} \,\,\,[\mbox{\cite{Inami}} ]& 
\sim 10^{-9} \,\,\,[\mbox{\cite{ElHRS}} ]
\\[0.5mm]
\hline
\end{array}
\end{eqnarray*}
\caption{Processes from which constraints on $Y_\nu$ and the $M_{N_i}$ 
could be obtained. The current and expected ex\-perimental sensitivity
are compared to the larger theoretical values obtained for various
textures of the matrix $H$. We didn't find this theoretical value 
for $d_\tau$
in the litterature but it is anyway much smaller than the 
experimental sensitivities. 
}
\end{table}
In Table.1 are shown the values of the various observables which can be reached
for some configurations of parameters satisfying these requirements. Also given
are the present and expected (in the forthcoming years) experimental 
sensitivities.
One observes that some of the experimental bounds (in particular 
$\mu \rightarrow e \gamma$ \cite{CaIb,ElILR,ElHRS}) can be 
already saturated for 
some configurations of the matrix $H$. With 
the experimental sensitivities expected 
in the future, a non-negligible fraction of the parameter space is 
expected to be covered, in particular for $\mu \rightarrow e \gamma$,
$\tau \rightarrow l \gamma$ and the EDM of the electron $d_e$.
Instead of considering chosen $H$ matrices, a 
more systematic way to scan the reachable theoretical values would be 
to take, in a random way, values of $\hat{c}_{1,2,3}$ in 
$R$, Eq.~(\ref{complexangles}).
Note also that the discovery of neutrinoless double beta decay
would provide one additional important constraint on the neutrino
mass matrix scale and Majorana phases. 
The discovery of supersymmetry and the measurements of the slepton masses
would be of course crucial in this program.
From all these observables one might get a chance to get 
constraints on $Y_\nu$ important for leptogenesis.
However, practically it's very unlikely that we are going to be in a 
corner of the parameter space for which several observables are expected 
to be observed soon. 
Moreover even if the matrix $H$ was determined with accuracy, the 
reconstruction of the $Y_\nu$ matrix from it still relies in this framework
on the 
assumption of universality of the soft mass terms at the grand unified 
scale. Would we relax this 
assumption, would we loose most of the model predictivity.
But waiting for more experimental information about this assumption 
it's definitely worth to proceed in this way!

\section{Extra low energy gauge boson and leptogenesis: the cases 
of $SO(10)$ and $E_6$}

There exist other types of low energy indications one might get on the
leptogenesis mechanism. If we embed the leptogenesis mechanism discussed above
in a grand unified theory (GUT), many low energy 
informations relevant for the GUT 
can also be relevant for leptogenesis.
Very important informations about the GUT group, the origin of neutrino 
masses and leptogenesis, would be furnished in particular 
by the discovery of an extra gauge
boson at low energy (around $\sim 1$~TeV). In the following we will 
discuss this possibility in the cases of $SO(10)$
and $E_6$ which are the two simplest groups which predict the existence
of singlet neutrinos in the same representation than the other fermions (i.e.
in the $16$
and $27$ representation 
respectively).
In the 
SO(10) case, the discovery 
of an extra $Z'$ or $W'$ basically would rule out the seesaw
mechanism of neutrino masses and leptogenesis with heavy singlet neutrinos.
This is due to the fact that there are no 
subgroup of $SO(10)$ larger than the SM group
which doesn't protect the $N_i$ masses from being large. For example
in the case where the left-right symmetry group 
$SU(3)_c \times SU(2)_L \times SU(2)_R \times U(1)_{B-L}$
would be broken 
at a low scale $\Lambda_{LR} \sim \mbox{TeV}$, since the $N_i$'s
are not singlets of this group they would acquire masses of the order 
of $\Lambda_{LR}$. This model can still be made consistent with the 
neutrino mass constraints, if the Yukawa couplings in $Y_\nu$ are taken 
sufficiently small in Eq.~(\ref{ooec}) but 
in this case a far too small asymmetry 
is produced in Eq.~(\ref{epsiloniresult}).\footnote{There 
exists an exception to this 
statement which consists in singlet neutrinos having a huge mass degeneracy
of order $10^{-10}$, in case the asymmetry is hugely enhanced 
by the terms in $1/(x_l-1)$ in Eq.~(\ref{epsiloniresult}) but 
we don't consider this 
possibility here.} Therefore to accommodate an extra low energy gauge boson
we need to consider 
the next anomaly free group which is $E_6$. The group $E_6$
is quite interesting because the discovery of an extra gauge 
boson is in this case compatible with
the neutrino mass and leptogenesis constraints, but only if this 
extra gauge boson is observed with given properties closely related
to these constraints\cite{E6}.
In fact since $E_6$ has rank 6  and the SM has rank 4, the SM can be extended
at low energy by an 
extra $U(1)$, $SU(2)$ or $SU(3)$. With a $SU(3)$ we could think 
about the $SU(3)_{R,L}$ of the maximal subgroup 
$SU(3)_c \times SU(3)_L \times SU(3)_R$ of $E_6$. With 
a $SU(2)$ there are three subgroups which correspond to
the three $SU(2)$ subgroups in $SU(3)_R$.
With a $U(1)$ there is a continuum of possibilities (i.e. any
combination of $U(1)_\psi$ and $U(1)_\chi$ which are defined by
$E_6 \rightarrow SO(10) \times U(1)_\psi$ and 
$SO(10) \rightarrow SU(5) \times U(1)_\chi$).
It turns out that among all these possible low energy extensions of the SM 
in $E_6$, only two have as singlets the $N_i$'s, that is to say 
allow the $N_i$'s to be heavy enough to satisfy
both the neutrino mass and leptogenesis constraints. These 
are \cite{E6}:
\begin{equation}
SU(3)_C \times SU(2)_L \times U(1)_Y \times U(1)_N \, ,
\end{equation}
\begin{equation}
SU(3)_C \times SU(2)_L \times SU(2)_R^A \times U(1)_{Y_L+Y'_R} \, .
\end{equation}
$U(1)_N$ is the combination of $U(1)_\psi$ and $U(1)_\chi$
for which the $N_i$'s are singlets (i.e. 
$Q_N= \cos \alpha Q_\psi + \sin \alpha
Q_\chi$ with $\tan \alpha = \sqrt{1/15}$)\cite{MaU1}. $SU(2)_R^A$ is
not the usual $SU(2)_R$ group of the left-right model
but the one\cite{MaSU2} in which $u_R$ is in a doublet with
$h_R$, the additional $SU(3)_c$ triplet in the $27$ fundamental
representation of $E_6$.

For these two subgroups one can show that the seesaw 
model works fine for leptogenesis
as well as for neutrino masses\cite{E6}. These two subgroups lead to a 
rich phenomenology at the TeV scale, in particular the presence
of 11 extra degrees of freedom at this scale in addition to the 
15 degrees of freedom of a full SM generation 
in each of the 3 generations of 27 multiplets.
It also predicts relations between the partial decay widths of the extra
gauge bosons to the various members of the 27 multiplets 
(i.e.~of the $Z'$ for $U(1)_N$ and the $Z'$ and the 2 $W'_R$
for $SU(2)_R^A$).
The observation at the LHC of this phenomenology, which is closely related 
to the neutrino mass and leptogenesis constraints, 
would provide strong indications for the seesaw model of neutrino 
masses and leptogenesis with heavy singlet neutrinos.

\section{Leptogenesis at the TeV scale}

So far we discussed the standard leptogenesis model based on the existence
of singlet
neutrinos with masses above $\sim 10^{10}$ GeV. We now want to address 
the question whether it is possible to build rather simple models at a directly
testable scale\cite{THTeV} (around $\sim 1$~TeV). In 
this case, instead of testing 
leptogenesis indirectly as discussed above, we could test it by producing
directly the particles at its origin.

\subsection{Problems occurring at the TeV scale}

We first discuss the three main problems one has to face if we want to build
such a low scale model:
\begin{itemize}
\item First, at such a low scale, the out-of-equilibrium condition for the
decay width, Eq.~(\ref{ooec}),
imposes the general condition that the couplings
are very tiny. This is due to the fact that first,
this condition is mediated by the very large
Planck scale and secondly, the decay width is in general only linear
in the mass of
the decaying particle, see Eq.~(\ref{ooec}), in contrast to the Hubble constant
which depends quadratically on
this mass.
This means that the product of couplings entering the decay width has to be
as much as 10 orders of magnitude smaller at the TeV scale than at
the $10^{13}$ GeV scale.
Beyond the fact that the naturality of such tiny couplings
can be questionable, the
major problem is that the associated produced asymmetry will be far too tiny,
due to the fact that the asymmetry
in most possible models is proportional to the same tiny couplings.
For example in the Fukugita-Yanagida model with singlet neutrinos 
having masses around
$\sim 1-10$ TeV, the produced asymmetry will be 
typically 6 orders of magnitude too small to account for $Y_B \sim 10^{-10}$.
\item At the TeV scale, various
scatterings can also be very fast with respect to the
Hubble constant. 
This is particularly the case
with gauge scatterings if the decaying particles producing the asymmetry
are not neutral or $SU(2)_L$ singlets. As shown in Ref.~\cite{THTeV} these 
scatterings easily wash out the asymmetry by 6 orders of magnitude. 
To avoid these effects,
the particle at the origin of the
asymmetry be better neutral and gauge singlet of any
low-energy gauge symmetry. This restricts largely the possibilities.
\item In the more ambitious and more interesting case where the source
of lepton number violation at the origin of the asymmetry is also at the
origin of the neutrino oscillations, an other problem could in
general occur. Two cases have to be distinguished.
First in the case where the
neutrino masses are produced at tree level, as
in the seesaw mechanism with singlet neutrinos,
the values
of the couplings which are needed to generate the neutrino masses are generally
slightly larger than
the ones allowed by the out-of-equilibrium condition.
At the 1-10 TeV scale this will 
induce an additional damping effect of order\cite{THTeV} $\sim 100$. 
Secondly in the case of neutrino masses generated by radiative
processes, as in the
Zee model\cite{Zee} or in R-parity violating supersymmetric
models\cite{Rparity}, it is quite difficult
to generate 
the neutrino masses without violating largely the out-of-equilibrium condition.
The
couplings necessary to accommodate the experimental neutrino mass constraints
are typically three order of magnitude larger than the values allowed by the
out-of-equilibrium condition for the associated decays. This induce 
a huge damping\cite{MaRaSa} of 
the produced 
asymmetry which exceeds $10^{6}$.

\end{itemize}

\subsection{A simple mechanism based on three body decays}

To avoid the three problems above one can think about three
different asymmetry enhancement mechanisms. The first is based on 
a huge mass degeneracy between at least two of the $N_i$'s \cite{Flanz} which
induces a resonant enhancement in Eq.~(\ref{epsiloniresult}) from the term
in $1/(x_l-1)$. For $M_{N_i} \sim 1$~TeV the 
degree of degeneracy required 
is of order $\Delta M_{N}/M_N \sim 10^{-10}$, which might be 
very difficult to test. For more explanations we refer 
the reader to Ref.~\cite{Flanz,Pil,BP,THTeV}.
A second mechanism consists in having a hierarchy among the couplings,
taking all the couplings tiny (to satisfy the out-of-equilibrium condition)
except the ones which intervene in the
one-loop decay amplitude but not in the tree level 
amplitude, i.e. in the numerator but not in the denominator 
of Eq.~(\ref{epsiloniresult}).
However, in the Fukugita-Yanagida model the neutrino constraints forbid 
this possibility \cite{THTeV,DaIb3}.
A third mechanism, which appears to our opinion to be more 
natural and definitely 
more testable, is based on three body decays\cite{THTeV}. If in the thermal
history of the universe the last L-violating
decay turns out to have been a three body decay, and not a two body 
decay as usually considered in usual leptogenesis models, and if it occurs at
a scale around $\sim 1-10$~TeV, then the lepton asymmetry produced 
can have the right order of magnitude without the need of any mass degeneracy.
This can be easily seen from the fact that in this case, for a three body decay
$A \rightarrow B+C^\ast \rightarrow B+D+E$, the numerator of the asymmetry
will involve 6 couplings and the numerator will involve 4 
couplings. In other words, if here for simplicity we take all 
these couplings to be equal ($\equiv g$), the asymmetry will be 
in $\sim g^2$ with a decay width in $\sim g^4$.
As a result, if the coupling $g$ is of order $\sim 10^{-3}$, the decay width
is suppressed enough to satisfy the condition $\Gamma < H$, 
with a produced asymmetry large enough (i.e. $\varepsilon \sim 10^{-8}$ which 
is perfectly fine).
Two body decays don't have this interesting property because they display
an asymmetry and a decay width which involve the same numbers 
of coupling (i.e. $\sim Y_\nu^2$ in Eqs.~(\ref{epsiloniresult}) 
and (\ref{ooec})).
Moreover a L-violating coupling $g$ of order $10^{-3}$ is typically 
what we need to induce at the TeV scale the neutrino masses in a radiative way
as in the Zee model and the models with R-parity violation.
This three body decay radiative mechanism 
may provide therefore a general framework of neutrino masses 
and leptogenesis at the TeV scale which is an alternative 
to the usual high scale
two-body decay seesaw framework. 

In Ref.~\cite{THTeV} we implemented 
this general framework with an explicit model. It 
is based on the three body decays of singlet neutrinos mediated by virtual
charged scalar singlets. 
The particle content of this minimal model consists in three 
singlet neutrinos $N_i$ 
having masses $\sim 1-10$~TeV, plus two charged scalar
singlets $S_{1,2}^+$ having similar 
masses (but heavier than the lightest singlet neutrino)
plus two lighter Higgs doublets $H_{1,2}$.
The tree level decays are represented 
in Fig.~1.
\begin{figure}[t]
\begin{center}
\begin{picture}(190,84)(0,0)
\Line(0,50)(33,50)
\ArrowLine(63,80)(33,50)
\DashArrowLine(33,50)(53,30){5}
\ArrowLine(53,30)(81,25)
\ArrowLine(53,30)(58,2)
\Text(5,42)[]{$N_i$}
\Text(63,68)[]{$l^c_j $}
\Text(52,46)[]{$S^-_{k} $}
\Text(81,32)[]{$l_m $}
\Text(65,3)[]{$l_n $}
\Line(105,50)(138,50)
\ArrowLine(168,80)(138,50)
\DashArrowLine(138,50)(158,30){5}
\DashArrowLine(158,30)(186,25){5}
\DashArrowLine(158,30)(163,2){5}
\Text(110,42)[]{$N_i$}
\Text(168,68)[]{$l^c_j $}
\Text(157,46)[]{$S^-_{k} $}
\Text(187,33)[]{$\phi_{1,2}^0 $}
\Text(174,4)[]{$\phi_{2,1}^- $}
\end{picture}
\end{center}
\caption{Singlet neutrino three body decays.}
\label{fig2}
\end{figure}
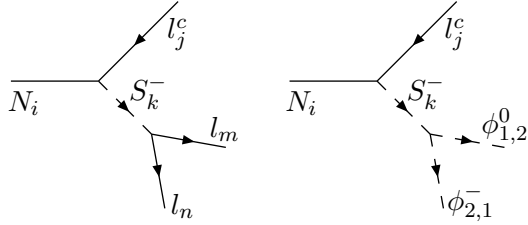
In this model the (Majorana) 
neutrino masses are generated as in the Zee model at 
the one loop level from the L-violating 
couplings of the charged scalar singlets
to two lepton doublets and to two
Higgs doublets. The leptogenesis is induced by the interference of the
tree level diagrams of Fig.~1 with one loop diagrams involving the 
self-energy of the 
virtual scalar singlets in Fig.~1 (with two leptons or two Higgs doublets
in the loop). 
In this model
the neutrino masses can also receive a contribution from the usual
Yukawa couplings in Eq.~(\ref{Lagr}). In order that this contribution
doesn't induce too 
large neutrino masses at this low scale through Eq.~(\ref{Mnu}), 
it is required 
in this model that these Yukawa couplings are small enough. This 
neutrino condition insures also that the usual 
two-body decays of the $N_i$'s don't over-dominate
the three body decays, otherwise this would suppress the asymmetry. 
More details can be found in Ref.~\cite{THTeV} where explicit
set of values of parameters reproducing the data for neutrino 
masses as well as leptogenesis are given.
This model shows what kind of minimal assumptions have to be made to build
a leptogenesis mechanism at a directly testable scale. The main assumption
is that a more involved particle content has to be considered.

\section*{Acknowledgments}
We thank E.~Ma, M.~Raidal and U.~Sarkar with whom part of the works 
presented here was done.
This work was
supported by the TMR, EC-contract No. ERBFMRX-CT980169 (EuroDa$\phi$ne).

\end{document}